\documentclass{arxiv_paper}

\pdfoutput=1

\usepackage{amsmath}
\usepackage{fancybox}
\usepackage{graphicx}
\usepackage{url}

\begin{document}

%
%
%


\large

\begin{titlepage}

   \date{25 May 2020}

   \title{Shape of Big Rockets}

  \begin{Authlist}
Thomas Hebbeker\Instfoot{iiia}{RWTH Aachen University, 
        Physics Institute III A}
  \end{Authlist}

  \begin{abstract}
    Recent publications
    discuss the size of
    chemical rockets for long-distance travel, to be launched from a planet.
    Here I point out that such a rocket cannot be tall and slim but will be short and fat.

  \end{abstract} 
  
\end{titlepage}

\setcounter{page}{2}

\section{Motivation}

Recently, M. Hippke
calculated the size a chemical rocket must have to leave the gravitational field of a planet
\cite{hippke1,hippke2}. He found that the rocket mass, 
given to a large extent by the fuel,
becomes enormous for a massive (exo)planet.
For simplicity
only single stage rockets are considered.
The author illustrates in Fig. 1 of reference~\cite{hippke1} how big such a rocket would be in comparison
to the Saturn V and other rockets that have been used on earth in the past, by scaling up their
linear dimensions, without changing the shape.

However, such a rocket has not only to be able to exceed the escape velocity of the planet and of the stellar
system it lives in, it must also 
be able to overcome locally the gravitational field at the surface of the planet. This requires
a huge boost during launch, which implies a wide rocket engine and thus a large diameter of the rocket.

\section{Calculation of Rocket Parameters}

We assume a planet with mass $M_P$ and radius $R_p$.
The single stage chemical rocket's mass is $m_R$ which is equal to the total fuel mass $m_F$ at the start,
thus we neglect the mass of the empty rocket and the payload. 
We denote by $v_F$ the exhaust velocity of the burnt fuel relative to
the rocket, which we assume to be constant. The mass of the expelled burnt fuel per time,
the mass flow rate 
$\dot{m_F} = \dot{m_R}$
should not change either, till the fuel is used up.
Atmospheric friction and other disturbances are not taken into account. 

In the launch position the rocket generates a lift off force
\begin{eqnarray}
  F_R
  = v_F \cdot \dot{m_R}  \; \; .
\end{eqnarray}
  This force $F_R$ must exceed the gravitational force $F_G$ between planet and rocket to make a lift off possible:
\begin{eqnarray}
  F_R > F_G
  \label{condition}
\end{eqnarray}
$F_G$ is the  rocket's weight at the surface of the planet, 
\begin{eqnarray}
  F_G = g_P \cdot m_R   \;\; , 
\end{eqnarray}
where
\begin{eqnarray}
  g_P  = G_N \cdot \frac{M_P}{R_p^2}
\end{eqnarray}
is the surface gravitational acceleration and $G_N$ denotes the gravitational constant. 
The condition
(\ref{condition}) 
translates into
\begin{eqnarray}
  \dot{m_R} > \frac{G_N \, M_P}{v_F \,  R_p^2} \cdot m_R
    \label{condition_explicit}
\end{eqnarray}
In fact, $F_R \gg F_G$
is desirable, else lots of fuel is burnt
just to
balance the gravitational pull during launch, but in the following estimate
we use the minimum requirement (\ref{condition}). 

Let's assume the rocket is of cylindrical shape with cross sectional area $A$ and height $H$,
so that the volume is $V = A \cdot H = m_R/\rho_R$ with the density of the unburnt fuel  $\rho_R = \rho_F$.
Ideally the burnt fuel can be expelled over the full area $A$, so the engine nozzle
which produces the exhaust jet covers the 
bottom of the rocket completely: 
\begin{eqnarray}
  \dot{m_R} =  v_F \cdot A \cdot \rho_E
  \label{mrdot}
\end{eqnarray}
where $\rho_E \ll \rho_F$ is the mass density of the exhaust gas. 
Putting equations (\ref{condition_explicit}) and (\ref{mrdot}) together yields
%
\begin{eqnarray}
  A   > \frac{G_N \, M_P}{v_F^2 \,  R_p^2 \, \rho_E} \cdot m_R
  = \frac{g_P}{v_F^2 \, \rho_E} \cdot m_R 
  \label{amin}
\end{eqnarray}
Thus the minimal cross section $A_{min}$
grows proportional to the rocket mass $m_R$, if all other model parameters are kept fixed. This implies that the corresponding height $H = m_R/(A \, \rho_F)$
is constant, and the ratio $H/A$ decreases, the rocket becomes fat.

\section{Numerical Examples}

Finally let's look at a simple numerical example, inspired by the
first stage of the Saturn V rocket as used for the first manned moon flights:
\begin{eqnarray*}
  g_P & = &  10 \, \mathrm{m/s^2}   = g_{Earth}  \\
%
  v_F & = & 3 \cdot 10^3 \, \mathrm{m/s} \\
%
  \rho_F & = & 10^3 \, \mathrm{kg/m^3} \\
%
  \rho_E & = & 0.1 \, \mathrm{kg/m^3} \\
%
  A & = & \pi \cdot (5 \, \mathrm{m})^2  = 80 \, \mathrm{m^2} \\  %
%
  H & = & 40 \, \mathrm{m} \\
%
  m_R &  = & 2 \cdot 10^6 \, \mathrm{kg}
\end{eqnarray*}
This gives for the smallest area $A_{min}$ fulfilling (\ref{amin}) 
\begin{eqnarray}
  A_{min} =  \frac{g_{Earth}}{v_F^2 \, \rho_E} \cdot m_R  = 20 \, \mathrm{m^2}
\end{eqnarray}
Not surprisingly, this is well below $A = 80 \, \mathrm{m^2}$. The main reason
for
$A_{min}/A < 1 $
is that
the Saturn V is designed such that $F_R$ is a factor of about three bigger than $F_G$,
in addition the total nozzle area is smaller than
the geometrical cross section $A$.

For a given ratio of $A_{min}/A$
a big rocket with $m_R = 4 \cdot 10^8 \, \mathrm{kg}$, as discussed in~\cite{hippke1,hippke2},
would have a cross section 200 times larger than the Saturn V, translating into a radius
of $70 \, \mathrm{m}$ instead of $5 \, \mathrm{m}$. The height would remain unchanged, thus
the diameter would exceed the height: we get a big fat rocket. 

%

\bibliographystyle{unsrt}  

\end{document}